\documentclass[aps,prd,twocolumn,superscriptaddress,nofootinbib,eqsecnum]{revtex4}

\pdfoutput=1

\usepackage{amsfonts}
\usepackage{amsmath}
\usepackage{amssymb}
\usepackage{graphicx,color}
\usepackage{float}
\usepackage{hyperref}
\usepackage{subfigure}
\usepackage{dcolumn}

\begin{document}

\title{Probability of Warm Inflation in Loop Quantum Cosmology}

\author{L. L. Graef} 
\affiliation{Departamento de F\'{\i}sica Te\'orica, Universidade do
Estado do Rio de Janeiro, 20550-013 Rio de Janeiro, RJ, Brazil}

\author{Rudnei O. Ramos} 
\affiliation{Departamento de F\'{\i}sica Te\'orica, Universidade do
Estado do Rio de Janeiro, 20550-013 Rio de Janeiro, RJ, Brazil}

\begin{abstract}

Warm inflation is analyzed in the context of Loop Quantum Cosmology
(LQC). The bounce in LQC provides a mean through which a Liouville
measure can be defined, which has  been used previously to
characterize the a priori probability for inflation in LQC. Here we
take advantage of the tools provided by LQC to study instead the  a
priori probability for warm inflation dynamics in the context of a
monomial quartic inflaton potential. We study not only the question of
how a general warm inflation dynamics can be realized in LQC with an
appropriate number of e-folds, but also how such dynamics is
constrained to be in agreement with the latest cosmic microwave
background radiation  from Planck. The fraction of warm inflation
trajectories in LQC that gives both the required minimum amount
e-folds of expansion and also passes through the observational window
of allowed values for the tensor-to-scalar ratio and the spectral tilt
is explicitly obtained. We find that the probability of warm inflation
with a monomial quartic potential in LQC is higher than that of cold
inflation in the same context. {}Furthermore, we also obtain  that
the a priori probability gets higher as the inherent dissipation of
the warm inflation dynamics increases.
 
\end{abstract}

\maketitle

\section{Introduction}
\label{intro}

The inflationary scenario~\cite{inflation} is the current paradigm for
the evolution of the early universe. Inflation is able to explain
several puzzles of the standard Big Bang cosmology, solving the
horizon and flatness problems and explaining the origin of inhomogeneities in the
universe, leading to large scale structure formation, based on causal
physics.  Despite of its great success, standard inflation has always
been challenged by several conceptual problems since its inception.  The need for
a large amount of fine-tuning in the parameters of the model, the
difficulties associated with the  process of reheating and the problem
in determining an appropriated measure with which one can determine
probabilities~\cite{Gibbons:2006pa}, are some of the problems which
has been topic of much
debate~\cite{Ijjas:2013vea,Guth:2013sya,Linde:2014nna}.  In addition,
after the Planck results~\cite{Planck2013,Ade:2015lrj}, some of the simplest
inflationary models, like the monomials quadratic and quartic
potentials for the inflaton field, were also disfavored by the data.

It is interesting to note that some of the aforementioned problems
have been analyzed in the context of warm inflation (WI)
dynamics. WI~\cite{Berera:1995ie} differentiates itself from the usual
cold inflation (CI) picture by accounting for the possible
nonequilibrium dissipative processes emerging from the microscopic
dynamics of the inflaton field and its necessary couplings to other
particle fields degrees of freedom. It has been shown that
nonequilibrium dissipation still during the slow-roll evolution of the
inflaton field can be potentially strong enough to produce a quasi
equilibrium  thermal radiation bath during inflation. Along with the
produced dissipation effect, this can strongly affect the  dynamics of
inflation both at the background and perturbation levels (for reviews
on WI, see, e.g.,  Refs.~\cite{Berera:2008ar,BasteroGil:2009ec}). WI
has much developed in recent years such as to be able to provide
insights on some outstanding problems related to the inflationary
picture and which CI cannot directly answer.  Some of these insights
include for example the question of eternal inflation in the context
of WI~\cite{Vicente:2015hga}, a solution for the so called
$\eta-$problem~\cite{Berera:1999ws} that appears in supergravity or
string inflation models, the role of dissipation in selecting
favorable inflationary trajectories and in alleviating the fine-tuning
problem of inflation~\cite{Ramos:2001zw,Berera:2000xz} (see also
Ref.~\cite{Bastero-Gil:2016mrl} for a recent detailed analysis of the
effect of dissipation and stochastic noise effects, typical of warm
inflation dynamics, in setting initial conditions for inflation).
This latter point is also connected with the naturalness of
inflation. It intends to address the question of whether a
sufficient duration of slow-roll dynamics can occur naturally or if it requires
a careful fine-tuning of initial parameters. Already in
Ref.~\cite{deOliveira:1997jt}  it was shown that inflation can be more
general in the presence of dissipation.  As it is well
known~\cite{Ramos:2013nsa,Bartrum:2013fia,Bastero-Gil:2014jsa,Panotopoulos:2015qwa,Visinelli:2016rhn}
WI is also able to decrease the tensor-to-scalar  ratio value,
rehabilitating several models ruled out in the CI context by the
Planck data.  In particular, recently in Ref.~\cite{Benetti:2016jhf}
it was analyzed how several primordial inflaton potentials can be
rendered compatible with the recent Planck data in the context of WI.

In spite of all the progress made in CI and WI, neither can address
one of the most important aspects of the Big Bang cosmology, the
singularity problem, which is in fact a problem related to the General
Relativity (GR).  {}For any equation of state obeying the strong
energy condition $p > - \rho/3$, regardless of the geometry  of the
universe, the scale factor in a
{}Friedmann-Lema\^itre-Robertson-Walker (FLRW) metric vanishes at $t =
0$, and the matter density diverges, leading to a collapse in FLRW
geometry. In fact, all the curvature invariants become singular, which
is the reason why this is called the Big Bang singularity problem.
However, it is common sense that in approaching the Big Bang
singularity we should enter in a regime where modifications of GR are
important, since ultraviolet effects  are  expected to become relevant
in the large curvature regime. It was shown that in some scenarios
these effects can smooth the Big Bang singularity.  In particular, one
of these scenarios, which is believed to be a possible candidate for a
quantum theory of gravity, is Loop Quantum Gravity (LQG) (see, e.g.,
Refs.~\cite{Bodendorfer:2016uat,Chiou:2014jwa} for some recent
reviews).  LQC arises as the result of applying principles of LQG to
cosmological settings  (see, e.g.,
Refs.~\cite{Ashtekar:2011ni,Barrau:2013ula,Agullo:2016tjh} for recent
reviews).  In LQC, which is the formalism  we will be focusing in this
work, the quantum geometry creates a brand new repulsive force that is
totally negligible at low space-time curvature but rises very rapidly
in the Planck regime, overwhelming the classical gravitational
attraction. In cosmological models, while Einstein's equations hold to
an excellent degree of approximation at low curvature, they undergo
major modifications in the Planck regime: {}For matter satisfying the
usual energy conditions any time a curvature invariant grows to the
Planck scale, quantum geometry effects dilute it, thereby resolving
singularities of GR~\cite{Ashtekar:2011ni,Ashtekar:2006rx,
  Ashtekar:2006wn, Ashtekar:2007em}.  In addition, as shown in
Refs.~\cite{Ashtekar:2009mm,Ashtekar:2011rm}, the ambiguity in
defining a measure present in GR   is naturally resolved in LQC
because in these scenarios the Big Bang is replaced by a Big Bounce,
and the bounce surface can be used to introduce the structure
necessary to specify a satisfactory Liouville measure. Using this
feature, the authors in Ref.~\cite{Ashtekar:2011rm} then makes a
detailed investigation of what would be the a priori probability of a
sufficiently long slow roll inflation in LQC which is compatible with
the data. Taking a monomial quadratic inflaton potential as an
example, they then showed that the a priori probability of having
enough inflation  (i.e., such that the number of e-foldings is $N_e
\gtrsim 67$ as considered in Ref.~\cite{Ashtekar:2011rm}) is very
close to one.  The probability problem of inflation in LQC was also
subsequently investigated in Ref.~\cite{Chen:2015yua}, while in
Ref.~\cite{Corichi:2010zp} a throughout interpretation of the
Liouville measure provided in LQC was given (see also
Refs.~\cite{Linsefors:2013cd,Bolliet:2017czc} for a different approach
for determining the probability of inflation in LQC).

In the present work,  we make a  detailed analysis of the dynamical
system of equations for inflation in LQC, including dissipative
effects in the context of WI.  By taking advantage of the fact that
LQC provides a  well defined measure with which one can compute
probabilities for  trajectories in phase space to have some minimum
number of e-folds, we then  apply that for WI. We firstly analyze the
fraction of  solutions in the phase space which gives  the expected
amount  of inflation, as done in  previous works in the context of
CI~\cite{Ashtekar:2011rm,Chen:2015yua}. Then we go a step further and
obtain also the fraction of those trajectories in WI that is
consistent with the Planck data, i.e., those trajectories that cross
the observational window of values for the tensor-to-scalar ratio $r$
and the spectral tilt $n_s$. This is done in particular for the
chaotic monomial quartic inflaton potential in the context of
WI. Recall that even though LQC can predict that inflation can be
quite natural with such potential, giving a probability quite close to
one for obtaining at least the minimum number of e-folds needed to solve the
flatness and horizon problems, the quartic potential in CI is well
excluded by the Planck data~\cite{Ade:2015lrj}.  Thus, even in LQC,
there would be simply a vanishing probability of rendering this
potential in CI compatible with the present observational data As
already mentioned, in WI the monomial quartic potential can be
rendered compatible with the observations  as a consequence of
dissipative effects. See in particular
Refs.~\cite{Bartrum:2013fia,Bastero-Gil:2016qru} for two particular
realizations in the context of WI  which we will  also consider in the
present work. Thus, we not only access the probability of WI in LQC,
but also its viability when enforcing that those inflationary
trajectories should necessarily satisfy the observations.  Although
there have been some previous studies of WI in
LQC~\cite{Herrera:2010yg,Xiao:2011mv,Zhang:2013yr,Herrera:2014mca,Basilakos:2017bol,Jawad:2017rkq,Kamali:2017zgg},
none of these works have focused on the questions we aim to answer in this
work and, in particular, on the probability of
WI in LQC. In addition, we also make use here of the most well
motivated forms of the dissipative dynamics in WI, as derived by
successful microscopic dynamics in nonequilibrium quantum field theory
applied to WI model building.

This paper is organized as follows. In Sec.~\ref{WILQC}, we briefly
introduce some of the key expressions in LQC and the respective
dynamics in the context of WI. Both the basic background quantities
and the results for the primordial power spectrum in WI are given.  In
Sec.~\ref{secphase}, we give some examples of evolution in the LQC
context and obtain the relevant input to be used in the computation of
the probability, based on the Liouville measure introduced in
Refs.~\cite{Ashtekar:2009mm,Ashtekar:2011rm}.  In Sec.~\ref{probab},
we compute the a priori probability of WI in LQC for different
dissipation models, focusing on the cases leading to consistency with
the Planck results. {}Finally, in Sec.~\ref{concl}, we give our
conclusions and comment on possible extensions of this work.

\section{Warm Inflation in LQC}
\label{WILQC}

Let us start this section by very briefly reviewing some of the
relevant results from LQC that will be useful for us. Then, we will
also present some of the equations and corresponding dynamics of WI.

\subsection{LQC dynamics}

In LQC cosmological models are quantized using the methods of
LQG. Below, we follow Ref.~\cite{Ashtekar:2011rm} in order to briefly
introduce the modification of the  {}Friedmann's equation in LQC.  The
spatial geometry in LQC is encoded in the volume of a fixed, fiducial
cubic cell, rather than the scale factor $a$, denoted by 
\begin{equation}
v = \frac{4 {\cal V}_0 a^{3} M_{\rm Pl}^2}{\gamma},
\label{eqv}
\end{equation}
where ${\cal V}_0$ is the comoving volume of the fiducial cell,
$\gamma$ is the Barbero-Immirzi parameter of LQC, whose numerical
value we set as given by $\gamma\simeq 0.2375$,  obtained from black
hole calculations~\cite{Meissner:2004ju}, and $M_{\rm Pl}\equiv
1/\sqrt{8 \pi G} = 2.4 \times 10^{18}$GeV is the reduced Planck
mass. The conjugate momentum to $v$ is denoted by $b$ and it is given
by
\begin{equation}
b=-\frac{\gamma P_{(a)}}{6 a^2 {\cal V}_0 M_{\rm Pl}^2},
\end{equation}
where $P_{(a)}$ is the conjugate momentum to the scale factor.  The
variables $v$ and $b$ satisfy the fundamental Poisson bracket,
$\{v,b\}=-2$.  The equation of motion for $v$ is given
by~\cite{Ashtekar:2011ni}
\begin{equation}
\dot{v} = \frac{3 v}{\gamma \lambda} \sin(\lambda b) \cos(\lambda b),
\label{dotv}
\end{equation}
where $\lambda$ is the constant
\begin{equation}
\lambda^2 = \frac{\sqrt{3} \gamma}{2 M_{\rm Pl}^2}.
\end{equation}
The solution of the LQC effective equations implies that the Hubble
parameter can be written as
\begin{equation}
H=\frac{1}{2 \gamma \lambda} \sin(2 \lambda b).
\label{eqH}
\end{equation}
In LQC, $b$ ranges over $(0, \pi/\lambda)$ and GR is recovered in the
limit $\lambda \rightarrow 0$.  The energy density $\rho$ relates to
the LQC variable $b$ through
\begin{equation}
\frac{\sin^2(\lambda b)}{\gamma^2 \lambda^2} = \frac{1}{3 M_{\rm
    Pl}^2} \rho,
\end{equation}
which combined with Eq.~(\ref{eqH}), leads to the {}Friedmann's
equation in LQC,
\begin{equation}
\frac{1}{9}\left(\frac{\dot{v}}{v}\right)^{2} \equiv H^2 = \frac{1}{3
  M_{\rm Pl}^2} \rho \left(1- \frac{\rho}{\rho_{\rm cr}} \right),
\label{Hubble}
\end{equation}
where 
\begin{equation}
\rho_{\rm cr} = \frac{3}{\gamma^2 \lambda^2} M_{\rm Pl}^2= \frac{2
  \sqrt{3}}{\gamma^3} M_{\rm Pl}^4.
\label{rhocr}
\end{equation}
We now see explicitly from Eq.~(\ref{Hubble}) that the singularity is
replaced by a (quantum) bounce when  $H=0$ and the density reaches the
critical value $\rho_{\rm cr}$. In particular, for $\rho \ll \rho_{\rm
  cr}$ we recover GR as expected. It is important to stress that the
expression~(\ref{Hubble}) holds independently of the particular
characteristics of the inflationary regime.

\subsection{WI background dynamics}
\label{WIcases}

In WI it is important  to account explicitly for the presence of
radiation. Hence, the total energy density in Eq.~(\ref{Hubble}) is
given by 
\begin{equation}
\rho = \frac{{\dot \phi}^2}{2} + V(\phi) + \rho_R,
\label{rho}
\end{equation}
which accounts for the presence of the radiation fluid and the scalar
field (the inflaton) $\phi$.  In this work we are going to consider
explicitly the monomial quartic potential for the inflaton
\begin{equation}
V(\phi) = \frac{\Lambda}{4} \left(\frac{\phi}{M_{\rm Pl}}\right)^4,
\label{Vphi}
\end{equation}
where $\Lambda/M_{\rm Pl}^4$ denotes here the (dimensionless) quartic
coupling constant.  The inflaton field $\phi$ and the radiation energy
density $\rho_R$ form a coupled system in WI dynamics, with background
evolution equations given, respectively, by
\begin{eqnarray}
&& \ddot \phi + 3 H \dot \phi + \Upsilon(\phi, T) \dot \phi +
  V_{,\phi}=0,
\label{eqphi}
\\ && \dot \rho_R + 4 H \rho_R = \Upsilon(\phi, T) \dot \phi^2,
\label{eqrhoR}
\end{eqnarray}
where $\Upsilon(\phi, T)$ is the dissipation coefficient in WI, which
can be a function of the temperature and/or the background inflaton
field. We will define below explicitly the two cases of dissipation
coefficient we will be considering in this work.  {}For a radiation
bath of relativistic particles, the radiation energy density is given
by  $\rho_R=\pi^2 g_* T^4/30$, where $g_*$ is the effective number of
light degrees of freedom  ($g_*$ is fixed according to  the
dissipation regime and interactions form used in WI).

The dissipation coefficient $\Upsilon$ in Eqs.~(\ref{eqphi}) and
(\ref{eqrhoR})  embodies the microscopic physics resulting from the
interactions between the inflaton and  the other fields that can be
present and accounts for the nonequilibrium dissipative processes
arising from these
interactions~\cite{Berera:2008ar,BasteroGil:2010pb}.  In particular,
we have two relevant cases of dissipation originating from previous
explicit quantum field theory model building for WI.     In the first
case, the inflaton is coupled to heavy intermediate fields, that are
in turn coupled to light radiation fields.  As the inflaton slowly
moves according to its potential, it can trigger the decay of these
heavy intermediate fields  into the light radiation fields  and
generates a dissipation term for the inflaton~\cite{Berera:2002sp}. In
this case, the resulting dissipation coefficient can be well described
by the
expression~\cite{Berera:2008ar,BasteroGil:2010pb,BasteroGil:2012cm}
\begin{equation}\label{upsilonT3}
\Upsilon_{\rm cubic} = C_{3} \frac{T^3}{\phi^2}, 
\end{equation}
where $C_{3}$ is a dimensionless parameter that depends on the
interactions specifics.   Hereafter we refer to the above
$\Upsilon_{\rm cubic}$ as the {\it cubic dissipation coefficient}.
This is obtained in the so-called {\it low temperature } regime for
WI~\cite{Berera:2008ar,BasteroGil:2010pb,BasteroGil:2012cm}, in which
the inflaton only couples to the heavy intermediate fields, whose
masses are larger than the radiation temperature and, thus, the
inflaton gets decoupled from the radiation fields. Typically, to be
able to generate large enough dissipation through the above mechanism,
it is required in general a large number of heavy intermediate
fields~\cite{BasteroGil:2012cm}.  This then reflects in the fact that
for relevant dynamics leading to WI in this case, the constant
$C_{3}$ in the Eq.~(\ref{upsilonT3}) is typically a very large
number. 

More recently, it was realized another mechanism able to lead to a
successful WI regime that requires minimal field
content~\cite{Bastero-Gil:2016qru}.  In
Ref.~\cite{Bastero-Gil:2016qru} an explicit WI model realization in
particle physics was constructed. It is based on a construction used
in Higgs phenomenology beyond the standard model, which uses a
collective symmetry where the inflaton is a pseudo-Goldstone boson.
In this case the inflaton can be directly coupled to the radiation
fields and gets protection from large thermal corrections due to the
symmetries obeyed by the model.  The resulting dissipative
coefficient, here obtained in the so called {\it large temperature}
regime (where the fields coupled to the inflaton are light with
respect to the ambient temperature), is given simply
by~\cite{Bastero-Gil:2016qru}
\begin{equation}\label{upsilonT1}
\Upsilon_{\rm linear} = C_{1} T,
\end{equation}
where $C_{1}$ is again a dimensionless parameter, like $C_3$ in
Eq.~(\ref{upsilonT3}), that  depends on the specific interactions of
the model and that leads to Eq.~(\ref{upsilonT1}) (see, e.g.,
Ref.~\cite{Bastero-Gil:2016qru} for details). In general we have that
$C_{1} \ll C_3$, reflecting the smaller field content required here as
compared with the cubic form of the dissipation coefficient.
Hereafter, we refer to the above equation~(\ref{upsilonT1}) as the
{\it linear dissipation coefficient}.

\subsection{The scalar curvature power spectrum in WI}

In WI not only the background dynamics gets modified, but also the
primordial power  spectrum can be strongly influenced by the presence
of the dissipative effects and the  produced radiation bath. WI at the
perturbation level has been studied by many works previously (see,
e.g.,
Refs.~\cite{Hall:2003zp,Graham:2009bf,BasteroGil:2011xd,Bastero-Gil:2014raa,Visinelli:2014qla}).
The power spectrum in WI can be written explicitly
as~\cite{Ramos:2013nsa} 
\begin{eqnarray} \label{spectrum}
\!\!\!\!\!\!\!\!\!\Delta_\mathcal{R}\! =\!\left(\!\frac{ H_{*}^2}{2
  \pi\dot{\phi}_*}\!\right)^2\!\!\left(\!1\!  +\!2n_*
\!+\!\frac{2\sqrt{3}\pi Q_*}{\sqrt{3\!+\!4\pi Q_*}}{T_*\over
  H_*}\!\right)\! G(Q_*),
\end{eqnarray}
where $Q_*$ denotes the ratio
\begin{equation}
Q_*= \frac{\Upsilon(T_*,\phi_*)}{3 H_*}.
\label{Qstar}
\end{equation}
The  subindex ``$*$" indicates that the quantities are evaluated at
Hubble radius  crossing.  The quantities in the primordial power
spectrum of Eq.~(\ref{spectrum}) are then evaluated when the relevant
CMB modes  cross the Hubble radius around $N_* \approx 50 - 60$
e-folds before the end of inflation. In this work we will consider
$N_*=60$ for definiteness.  In Eq.~(\ref{spectrum}), $n_*$ denotes the
inflaton statistical distribution due to the presence of the radiation
bath.   Here, as also in previous works (e.g., like in
Ref.~\cite{Benetti:2016jhf}), we will assume a thermal equilibrium
distribution function $n_* \equiv n_{k_*}$ for the inflaton and, thus,
it assumes the Bose-Einstein distribution form, $n_* =1/[\exp(H_*/T_*)
  -1]$. In fact, as shown recently in Ref.~\cite{Bastero-Gil:2017yzb},
due to the presence of the radiation bath and the inflationary
dynamics itself in WI, there can always be a nonvanishing  statistical
distribution for the inflaton  which can be close to the thermal one,
even if the inflaton interaction rate with the radiation bath
particles is smaller than $H$.  The function $G(Q_*)$ in
Eq.~(\ref{spectrum}) accounts for the growth of inflaton fluctuations
due to its coupling with the radiation fluid (this comes explicitly
from the expression for the dissipation coefficient $\Upsilon$, which
in general is an explicit function of the temperature, thus coupling
both inflaton and radiation perturbations). It  can only be determined
numerically by solving the full set of perturbation equations found in
WI~\cite{Graham:2009bf,BasteroGil:2011xd,Bastero-Gil:2014jsa,Bastero-Gil:2016qru}.
According to the method of the previous works, we use a numerical fit
for $G(Q_*)$.  Here we follow the notation used in
Ref.~\cite{Benetti:2016jhf} and we consider for the linear dissipation
coefficient $\Upsilon_{\rm linear}$, that $G(Q_*)$ is given by
\begin{eqnarray} \label{growing_mode}
G_{\rm linear}(Q_*)\simeq 1+ 0.335 Q_*^{1.364}+ 0.0185Q_*^{2.315},
\end{eqnarray} 
while for the cubic dissipation coefficient,  $\Upsilon_{\rm cubic}$,
$G(Q_*)$ is given by
\begin{eqnarray} \label{growing_mode2}
G_{\rm cubic}(Q_*)\simeq 1+ 4.981 Q_*^{1.946}+ 0.127 Q_*^{4.330}.
\end{eqnarray} 
Note that the CMB data constrains the amplitude of the scalar
curvature power spectrum at  a pivot scale $k_*$ as being
$\Delta_{{\cal R}}(k=k_*) \simeq 2.2 \times 10^{-9}$, with $k_*=0.05
{\rm Mpc}^{-1}$, as considered by the Planck
Collaboration~\cite{Ade:2015lrj}. This is the CMB normalization we use
in this work.

Given the scalar curvature power spectrum expression,
Eq.~(\ref{spectrum}),  the tensor-to-scalar ratio $r$ and the spectral
tilt $n_s$ follow from their usual definitions, just like in the CI
case,
\begin{equation}
r= \frac{\Delta_{T}}{\Delta_{{\cal R}}},
\label{eq:r}
\end{equation}
and
\begin{equation}
n_s -1 = \lim_{k\to k_*}   \frac{d \ln \Delta_{{\cal R}}(k/k_*) }{d
  \ln(k/k_*) },
\label{eq:n}
\end{equation}
where $\Delta_{T} = 2 H_*^2/(\pi^2 M_p^2)$ is the tensor power
spectrum.  Due to the weakness of gravitational interactions, the
tensor modes are expected not to be affected by the dissipative
dynamics and $\Delta_{T}$ remains essentially unaltered from the CI
result~\cite{Ramos:2013nsa} (see also Ref.~\cite{Li:2018wno} for a
recent study on the possible changes of the tensor spectrum in WI).

\begin{center}
\begin{figure}[!htb]
\includegraphics[width=8cm]{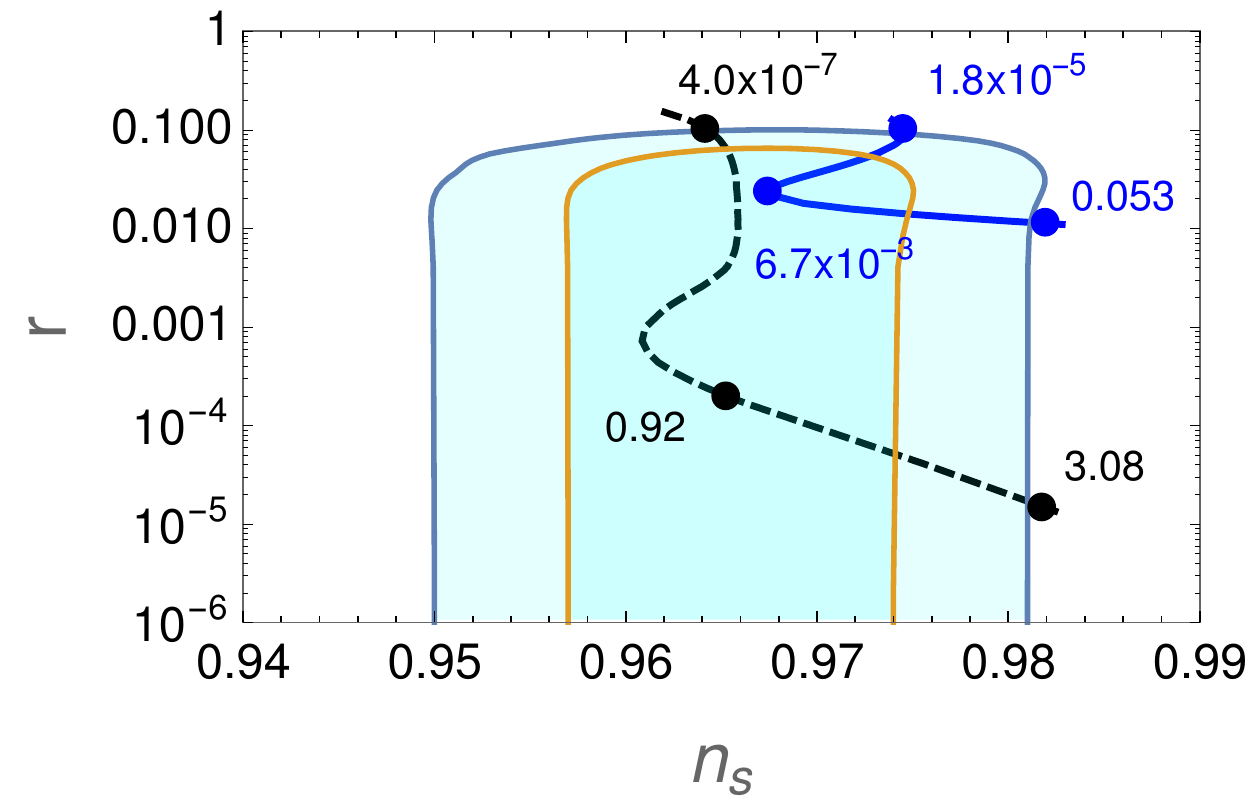}
\caption{The spectral index $n_s$ and the tensor-to-scalar ratio $r$
  in the plane $(n_s,r)$ for different values of the dissipation ratio
  $Q_*$ (indicated by the numbers next to the curves), for the chaotic
  quartic potential. The blue solid curve shows the case of a
  dissipation coefficient which is  cubic in the temperature while the
  black dashed curve shows the case of a dissipation coefficient which
  is linear in the temperature.    The contours are for the $68\%$ and
  $95\%$ C.L. results from Planck 2015 (TT+LowP+BKP data). }
\label{fignsXr}
\end{figure}
\end{center}

In {}Fig.~\ref{fignsXr} we show the results obtained for both $r$ and
$n_s$ in WI for the two forms of dissipative coefficients  given by
Eqs.~(\ref{upsilonT3}) and (\ref{upsilonT1}). {}For reference, in each
dissipation case, we have indicated the values of the dissipation
ratio $Q_*$ for which we can find consistency with the Planck results.
{}For the cubic dissipation coefficient form, we thus find that WI
with the quartic inflaton potential can be rendered  compatible with
the Planck data for values of dissipation ratio $Q_*$ satisfying  $1.8
\times 10^{-5} \lesssim Q_* \lesssim 0.053$. {}For the case with the
linear dissipation coefficient,  even a larger range of values is
allowed, given by $4.0 \times 10^{-7} \lesssim Q_* \lesssim 3.08$.
Note that in CI, for the quartic potential we have that $r\simeq
0.266$ and $n_s \simeq 0.95$, thus laying well outside the allowed
Planck region. WI dissipative effects effectively decreases the
tensor-to-scalar ratio, making the quartic potential compatible with
the data (see also Refs.~\cite{Bastero-Gil:2017wwl,Arya:2017zlb} for
recent works showing the compatibility of the quartic potential in WI
with the Planck data).

\subsection{Corrections to the power spectrum from LQC}
\label{effects}

To get the results given in the previous subsection, we have
considered the dynamics as in the standard GR case, thus neglecting
the LQC correction to the {}Friedmann's equation (\ref{Hubble}). This
is consistent if we compare the total energy densities in  the cases
studied here with the critical density Eq.~(\ref{rhocr}).  {}For
instance,  for the cases shown in {}Fig.~\ref{fignsXr}, we have that
the largest energy density for the cubic dissipation case  is reached
for the value $Q_*=1.8 \times 10^{-5}$. The energy density in this
case is  $\rho_* \equiv \rho_{\phi,*} + \rho_{R,*} \simeq 3.45 \times
10^{-9} M_{\rm Pl}^4$. {}For the linear  dissipation case this happens
at $Q_*= 4.0 \times 10^{-7}$, with an energy density $\rho_* \simeq
3.39 \times 10^{-9} M_{\rm Pl}^4$.  Hence, in both cases we have that
$\rho_* \ll \rho_{\rm cr} \simeq 258.58 M_{\rm Pl}^4$ and quantum
effects on the geometry from LQC can be neglected in principle.
However, even though at the background level this can be certain, we
have to evaluate with care the LQC contributions at the perturbation
level. In particular, the power spectrum can receive important
contributions due to LQC, including even the presence of features, as
recent works have shown.

In the cold inflationary scenarios in GR, it is assumed that the
pre-inflationary dynamics does not have any effect on modes that are
observable in the CMB. In these scenarios, these modes  have physical
wavelengths smaller than the curvature radius at the onset of
inflation and all the way back to the Big Bang. Modes with such
wavelengths do not feel the effects of the curvature, and then
propagates as if they were in flat spacetime, with a trivial
dynamics. However, the situation is different in LQC due to the
distinct pre-inflationary dynamics. In this scenario some of the
relevant modes have physical wavelengths comparable to the curvature
radius at the bounce time. As showed by Parker~\cite{parker}, modes
that experience curvature are excited. So, large wavelength modes are
excited in the Planck regime that follows the bounce. As a
consequence, at the onset of inflation, the quantum state of
perturbations is populated by excitations of these modes over the
Bunch-Davis vacuum, changing the initial conditions for perturbations
at the onset of inflation~\cite{agulloetal}.  Due to this, the scalar
curvature power spectrum in LQC gets modified with respect to GR, such
that it can be written as\footnote{Here, we will follow the notation
  of Ref.~\cite{Zhu:2017jew}, where a complete derivation of these
  pre-inflationary LQC modifications to the GR spectrum were given.}
\begin{equation}\label{PR1}
\Delta_\mathcal{R}(k)= |\alpha_{k} +\beta_{k}|^{2}
\Delta_\mathcal{R}^{GR}(k),
\end{equation}
where $\alpha_{k}$ and $\beta_{k}$ are the Bogoliubov coefficients,
where the pre-inflationary effects are codified  and
$\Delta_\mathcal{R}^{GR}$ is the GR form for the power spectrum, e.g.,
given by Eq.~(\ref{spectrum}).  Below, we follow, in particular, the
derivation done recently in  Ref.~\cite{Zhu:2017jew}.

Equation~(\ref{PR1}) can be parametrized as
\begin{equation}\label{PR}
\Delta_\mathcal{R}(k)= (1+\delta_{PL})\Delta_\mathcal{R}^{GR}(k).
\end{equation}
The factor $\delta_{PL}$ in the above equation is k-dependent and it
takes into account the LQC corrections. It is given by~\cite{Zhu:2017jew}
\begin{align}\label{complete}
&\delta_{PL} =  \left[1+\cos\left(\frac{\pi}{\sqrt{3}}\right)\right]
  {\rm csch}^{2}\left(\frac{\pi k}{\sqrt{6} k_{B}}\right)   \nonumber
  \\ &+ \sqrt{2} \sqrt{\cosh\left(\frac{2\pi k}{\sqrt{6}
      k_{B}}\right)+\cos\left(\frac{\pi}{\sqrt{3}}\right)}
      \cos\left(\frac{\pi}{2\sqrt{3}}\right)
  \nonumber  \\ &\times {\rm csch}^{2}\left(\frac{\pi
    k}{\sqrt{6}k_{B}}\right)\cos(2k \eta_{B}+\varphi_{k}),
\end{align}
where $\varphi_{k}$ is given by
\begin{equation}\label{short}
\varphi_{k} \equiv \arctan \left\{\frac{{\rm
    Im}[\Gamma(a_{1})\Gamma(a_{2})\Gamma^{2}(a_{3} - a_{1} - a_{2})]}
       {{\rm Re}[\Gamma(a_{1})\Gamma(a_{2})\Gamma^{2}(a_{3} - a_{1} -
           a_{2})]}\right\},
\end{equation}
where the $a_1,\,a_2,\, a_3$ are defined as $a_{1,2} = (1\pm
1/\sqrt{3})/2 -i k/(\sqrt{6} k_B)$ and $a_3=1-ik/(\sqrt{6} k_B)$ and
the index $B$ in the quantities in the above equations indicates that
they are calculated at the bounce.  In particular, $\eta_B$ is the
conformal time at the bounce and $k_B= \sqrt{\rho_{\rm cr}} a_B/M_{\rm Pl}$
is a characteristic  scale at the bounce.

The term in Eq.~(\ref{complete}) with $\cos(2k \eta_{B}+\varphi_{k})$
oscillates very fast and has negligible effect when averaging out in
time.  The factor $\delta_{PL}$ then simplifies to
\begin{equation} \label{delta}
\delta_{PL} = \left[1+\cos\left(\frac{\pi}{\sqrt{3}}\right)\right]
      {\rm csch}^{2}\left(\frac{\pi k}{\sqrt{6}k_{B}}\right).
\end{equation}
Since this pre-factor represents the effects of the pre-inflationary
dynamics, it has the same expression in both the cold and
warm inflationary pictures.

Likewise, the tensor spectrum in the LQC can be written
as~\cite{Zhu:2017jew}
\begin{equation}\label{PH}
\Delta_{T}(k)= (1+\delta_{PL})\Delta_{T}^{GR}(k),
\end{equation}
where $\Delta_{T}^{GR}(k)$ is the tensor spectrum in standard CI. As
already explained in the previous subsection, due to the weakness of
gravitational interactions, $\Delta_{T}^{GR}(k)$ in WI has basically the
same expression as in CI.  It is remarkable to note that the
pre-factor $\delta_{PL}$ for both scalar and tensor perturbations are
equal, and are  given by Eq.~(\ref{delta}). As a result, the
tensor-to-scalar ratio is the same as that given in GR.

As discussed by the authors of Ref.~\cite{Zhu:2017jew}, in standard
inflation, in addition to the usual number of e-folds $N_{*} \equiv
ln(a_{end}/a_{*}) \approx 60$ required in general, it is necessary an
extra amount of expansion $\delta N \equiv ln (a_{*}/a_{B}) \gtrsim 21$, in order
for the model to be consistent with observations. As shown in
Ref.~\cite{Zhu:2017jew}, after the effects of the pre-inflationary
dynamics are taken into account, the power spectra  are generically
scale-dependent and exhibits oscillatory features (note that
oscillatory features in the power spectrum in bouncing models is a
generic result, as shown, e.g., recently in
Ref.~\cite{Brandenberger:2017pjz}).  As a consequence, in order to be
consistent with observations, the universe must have expanded at least
21 e-folds from the bounce till Hubble radius crossing at the
observables scales, such as to allow for these scale-dependent
features to get sufficiently diluted away and not spoiling the
perturbation spectra of CMB.  This is why, in the analysis that will
follow below, in addition of the usual minimum 60 e-folds of inflation
(referring  to the amount of expansion from Hubble radius crossing
$N_*$ to the end of inflation), we will also consider at least 21
additional e-folds from the bouncing time till $N_*$. Therefore, we
will consider a total of at least 81 e-folds from the bounce till the
end of inflation, such  that any pre-inflationary effects are
sufficiently diluted and do not  spoil the observed spectrum
(see also Ref.~\cite{Wilson-Ewing:2016yan} and references therein for
a discussion about these and other LQC effects).

\section{Illustration of the dynamics in LQC for the CI and WI pictures}
\label{secphase}

\subsection{The phase space system}

It is always useful to look the background dynamics as a dynamical
system. In this way, the inflaton's equation of motion,
Eq.~(\ref{eqphi}), together with $\dot \phi = d \phi/dt$ and the time
derivative for the Hubble parameter,
\begin{equation}
\dot H \!= \!-\frac{1}{2 M_{\rm Pl}^2} \left( \dot \phi^2 \!+\!
\frac{4}{3} \rho_R \right)  \left[ 1\! - \! \frac{ \dot \phi^2 \! + \!
    2 V(\phi) \! + \! 2 \rho_R}{\rho_{\rm cr}} \right],
\label{eqH2}
\end{equation}
form a three-dimensional (dissipative) dynamical system\footnote{Note
  that $\rho_R$ can be expressed in terms of $\phi,\dot \phi$ and $H$
  from the equation for the Hubble parameter, Eq.~(\ref{Hubble}), and,
  hence,  can be considered as the first integral of
  Eq.~(\ref{eqrhoR}). Thus, there is no need to consider $\rho_R$ as
  part of the dynamical system.} in the phase space $(\phi,\dot \phi,
H)$. Since the radiation energy density satisfies $\rho_R\geq 0$ by
definition, we thus have, from Eq.~(\ref{Hubble}), that $\phi,\dot
\phi$ and $H$ satisfy the inequality
\begin{equation}
H^2 \geq  \frac{1}{3 M_{\rm Pl}^2} \left[ \frac{{\dot \phi}^2}{2} +
  V(\phi) \right] \left[1- \frac{\frac{{\dot \phi}^2}{2} +
    V(\phi)}{\rho_{\rm cr}} \right].
\label{eqphase}
\end{equation}
Hence, it defines a volume in the phase space $(\phi,\dot \phi, H)$
such that the physical region  $\rho_R\geq 0$ is inside this
volume. {}In LQC, the upper half of this volume, $H \geq 0$,
corresponds to those trajectories that evolve from the bounce time to
the future. In the absence of radiation, $\rho_R=0$, the equality in
Eq.~(\ref{eqphase}) represents a surface, or more particularly in the
language of dynamical systems, an {\it invariant manifold}.   As a
consequence, all initial conditions chosen on this surface will
produce trajectories that evolve on it and end towards the origin. While
for the dissipative system, in which $\rho_R \neq 0$, all trajectories with 
initial conditions taken inside this volume, they will
evolve inside the volume of the invariant manifold. Trajectories with initial
conditions taken externally to the invariant manifold, will evolve towards it,
being this region then an attractor in phase space.  The inflationary
region in this phase space is defined by the condition,
\begin{equation}
\frac{\ddot a}{a} = \dot H + H^2 >0.
\label{accel}
\end{equation}

\begin{center}
\begin{figure}[!htb]
\includegraphics[width=8cm]{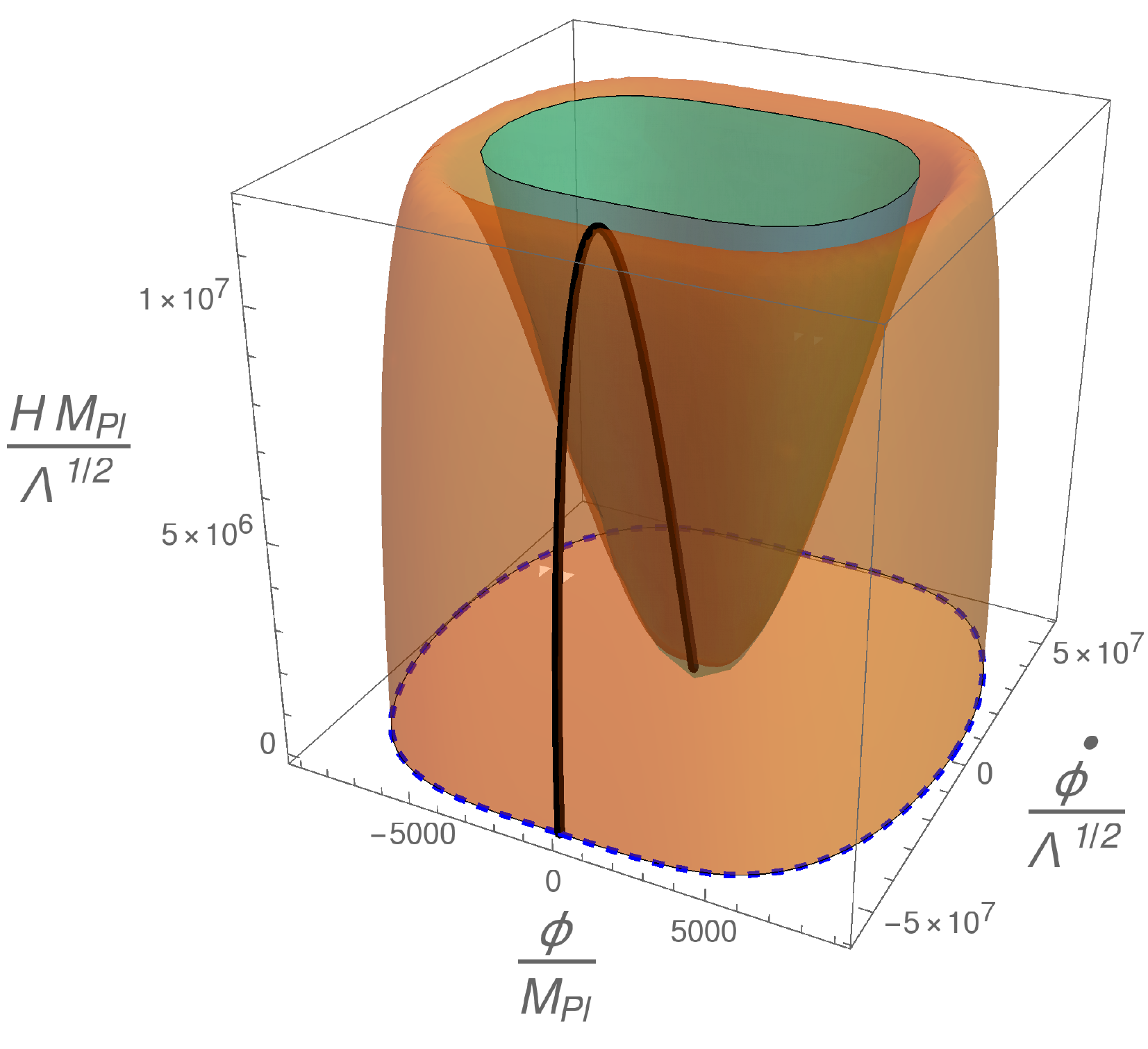}
\caption{The phase space available in the cases of LQC (external orange region)
  and in GR (internal green region).  The black curve shows an example of a
  trajectory evolving from the bounce forward in time. The horizontal dashed
  curve indicates the location of the bounce, in the plane $(\phi, \dot \phi)$.}
\label{figphasespace}
\end{figure}
\end{center}

In {}Fig.~\ref{figphasespace} we show an example of phase space volume
in LQC for the case of the quartic potential,
Eq.~(\ref{Vphi}), with a fixed value of coupling constant
$\Lambda/M_{\rm Pl}^4 \simeq 1.51\times 10^{-13}$.  {}For comparison,
we also show the corresponding volume in the GR case. Note that in
LQC, $H$ is naturally bounded from above, $H_{\rm max} = 1/(2 \gamma
\lambda)$, as seen explicitly from Eq.~(\ref{eqH}),  while, of course,
there is no such upper bound in GR. In {}Fig.~\ref{figphasespace} we
have truncated  $H$ in the GR case at the same value of $H_{\rm max}$ in LQC for
illustration. The two regions coincide when $\rho \ll \rho_{\rm cr}$,
but they departure from each other as we approach the critical density
value $\rho_{\rm cr}$. The bounce points (shown by the horizontal dashed curve) 
in the  phase space are clear in {}Fig.~\ref{figphasespace}. The invariant manifold is
always given by the external surface of the volumes in
{}Fig.~\ref{figphasespace}.  The black line in
{}Fig.~\ref{figphasespace} shows an example of trajectory in the phase
space in the CI case, for the quartic coupling value given above,
evolving from the bounce (with kinetic energy dominated initial
conditions) forward in time. We can see that the Hubble parameter
starts from $H=0$ at the bounce and then increases  until it reaches its
maximum value, when it starts decreasing again, reaching a small
positive value.  Since it is a case in the absence of dissipation
(CI), the trajectory evolves on the invariant manifold.

\subsection{Dynamics of WI in LQC}

Let us now expose some of the explicit results concerning the dynamics
evolution of WI in LQC. As already discussed in Sec.~\ref{intro}, our
interest is in investigating those cases for which the inflationary
trajectories will cross the favorable Planck region for the
observables $r$ and $n_s$. Therefore, we focus on the results leading
to those trajectories shown in  {}Fig.~\ref{fignsXr}. Each of the
indicated dissipative points shown in {}Fig.~\ref{fignsXr} can be
regarded as a different model, whose more fundamental origin can be
seen as coming from some explicit model building for WI, which we have
discussed in Sec.~\ref{WIcases} (but whose detailed particle physics
specifics we will not be concerned here).  Thus, we specify six cases
we are going to analyze, corresponding to the points shown in
{}Fig.~\ref{fignsXr}: (a) two initial conditions leading to
trajectories passing at the top extrema of the 2$\sigma$ CL of the
Planck contour, corresponding to the lowest values of the dissipation
(one for each of the two dissipation cases given previously); (b)  two
initial conditions (again, one for each dissipation coefficient form)
leading to trajectories passing at the center values of $(r,n_s)$ in
the Planck contour; and finally, (c) two initial conditions leading to
trajectories leaving the right extremum of the 2$\sigma$ CL of the
Planck contour (corresponding to the cases with the largest values of
dissipation coefficient consistent with the observational data).

Our numerical strategy is the following: {}First, the values of the
field, its time derivative and the radiation energy density are
obtained in each of these cases, i.e., the values of $\phi_* \equiv
\phi(N_*=60), \; \dot \phi_* \equiv \dot \phi(N_*=60)$ and $\rho_{R,*}
\equiv \rho_R(N_*=60)$; Then, we extrapolate the dynamics backwards in
time up to the bounce time (which from now on is assumed to be our
initial time, with $N_e=0$) and obtain the respective values of the field, its time
derivative and the radiation energy density at the bounce in each
case, $\phi_B, \; \dot \phi_B$ and $\rho_{R,B}$.  By doing this we are
certain that the evolution from this instant forward in time will lead
to trajectories passing through each of the points shown in
{}Fig.~\ref{fignsXr}.   As already mentioned, we focus on the cases in
which from the bounce initial time till the inflationary region (with
Hubble radius exit at $N_{*}=60$), there is at least 21 additional
e-folds of evolution, such as to allow enough time to dilute the
effects discussed in  Sec.~\ref{effects}. In addition,  we will focus
essentially in the case of a kinetic energy dominated bounce, i.e.,
$\rho_B \sim \rho_{\dot \phi,B} \equiv \dot \phi_B^2/2 \sim \rho_{\rm
  cr}$, with a negligible initial radiation energy
density\footnote{Note that here we assume that the radiation is
  produced by the inflaton dynamics from the bounce instant
  onwards. We are also not concerned in this work with any possible
  dynamics that might be present prior to the bounce, i.e., in the
  contracting phase of the universe.}. Thus, at the bouce, we always
have $\rho_{\dot \phi,B} \gg V(\phi_B) \gg \rho_{{\rm R},B}$.  We make
this choice since this is expected to be the most relevant case
concerning the conditions emerging from the bounce (see, e.g.,
Refs.~\cite{Ashtekar:2009mm,Ashtekar:2011rm}).  A bounce dominated by
the potential energy of the inflaton will always lead to extremely
large  number of e-folds of inflation (thus inflation is certain). In
the case of a radiation energy dominated bounce, besides not being a
common assumption in bouncing models (see, however,
Ref.~\cite{Cai:2014jla}), it also  prevents inflation in general,
except in those cases in which  the inflaton potential energy density
is large enough such that, as the radiation energy density dilutes
faster, $V(\phi)$ can come to dominate at some point and leading to
inflation (see, e.g., Ref.~\cite{Bastero-Gil:2016mrl} for a discussion
of the initial condition problem for inflation in this context).
Since this later case is rather model dependent,  we avoid addressing
such situation here.

\begin{table*}[!htb]
\caption{The values of parameters used to compute the probabilities.}
\begin{center}
\begin{tabular}{c|c|c|c|c|c|c|c}
  \hline  \hline 
  ${\rm Case}$ & $\Upsilon$ &  $Q_*$  & $C_i$  &  $\Lambda/M_{\rm Pl}^4$  & $\phi_B/M_{\rm Pl}$ & $\phi_{\rm min}/M_{\rm Pl}$ & $\phi_{\rm max}/M_{\rm Pl}$ 
  \\ 
\hline 
CI  & -   & -  &  - & $1.5115 \times 10^{-13}$  & $\pm 9095.1571$ & $-34.3917$   & $13.9391$
\\ 
\hline
  &   &  $1.8 \times 10^{-5}$ & $4.4772 \times 10^6 $  & $9.7785 \times 10^{-14}$    & $\pm 10141.3811$   & $-21.6089$   & $21.6089$
\\   
WI &  $\propto\frac{T^3}{\phi^2}$  &  $6.7 \times 10^{-3}$ & $1.1975 \times 10^{7}$  & $5.7204 \times  10^{-14}$ & $\pm 11595.9834$  & $-18.3440$  & $18.3440$
\\    
  &   & $0.053$ & $1.6377 \times 10^{7}$ & $4.6832 \times 10^{-14}$  & $\pm 12190.7109$ & $-16.6455$  & $16.6455$
\\ 
\hline 
   &  &  $4.0 \times 10^{-7}$ & $1.1561 \times 10^{-6}$ & $5.5389 \times 10^{-14}$  &  $\pm 11689.8322$ & $-35.8902$  & $13.4930$
\\  
WI  & $\propto T$  &  $0.92$ & $0.0121$ &   $1.6506\times 10^{-15}$ & $\pm 28135.362$  & $-22.1514$  & $5.5982$
\\    
  &  &  $3.08$ & $0.0181$ & $8.0446 \times 10^{-16}$ & $\pm 33673.5327$  & $-16.9156$ & $1.4951$
\\    
\hline
\end{tabular}
\end{center}
\label{tab:models} 
\end{table*}

The relevant values of parameters following from the above described
strategy are given in Tab.~\ref{tab:models}.  Since all previous
results in CI have focused essentially in the monomial quadratic
potential, for completeness we also give the results of CI for the
quartic potential, which will be useful for future comparison with
WI.  In Tab.~\ref{tab:models} we give the results for different
inflaton coupling constants $\Lambda/M_{\rm Pl}^4$ for each of the
cases studied. These values enforce that the amplitude of the
primordial scalar power spectrum is normalized with the Planck CMB
results, i.e., $\Delta_{{\cal R}}(k=k_*) \simeq 2.2 \times
10^{-9}$. The values of the dissipation coefficient $Q$ refers to
those given  in {}Fig.~\ref{fignsXr}. The values of $\phi_{\rm min}$
and $\phi_{\rm max}$ give the range of initial values of  inflaton
amplitude for which we have less that $N_e \simeq 81$ e-folds of
expansion from the bounce till the end of inflation, i.e., for field
values satisfying $\phi_{\rm min} \lesssim \phi \lesssim \phi_{\rm
  max}$, we have   $N_e \lesssim 81$ e-folds of expansion. The values
of $\phi_B$ given in Tab.~\ref{tab:models} are the maximum allowed
initial values of the inflaton field, which correspond to the cases of
a bounce  dominated by the potential energy $V(\phi)$. These values
are used in the definition of the probability that we are going to
compute in the next section following the method given by the authors
in    Refs.~\cite{Ashtekar:2009mm,Ashtekar:2011rm}.

\begin{center}
\begin{figure}[!htb]
\subfigure[CI]{\includegraphics[width=6cm]{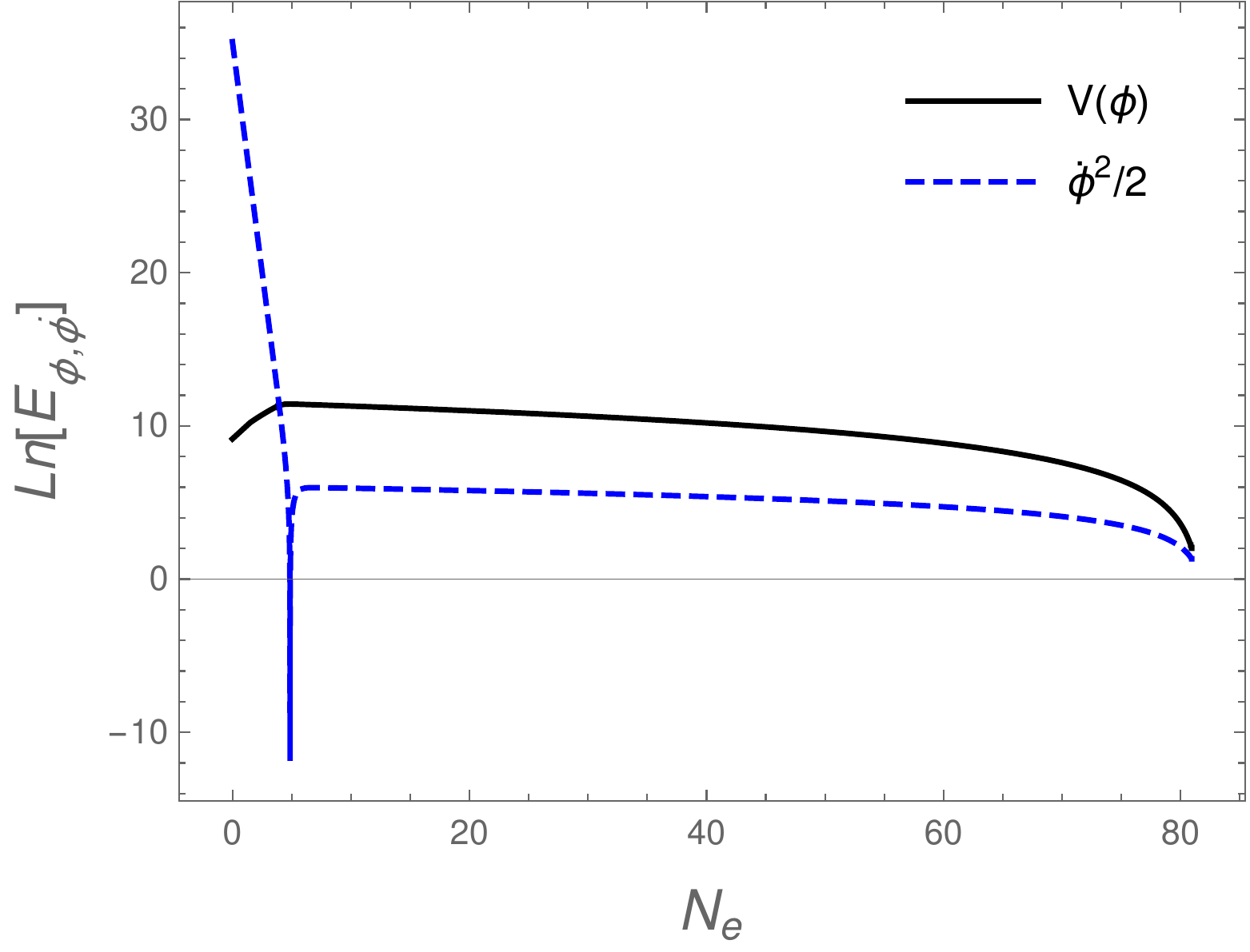}}
\subfigure[WI: cubic
  dissipation]{\includegraphics[width=6cm]{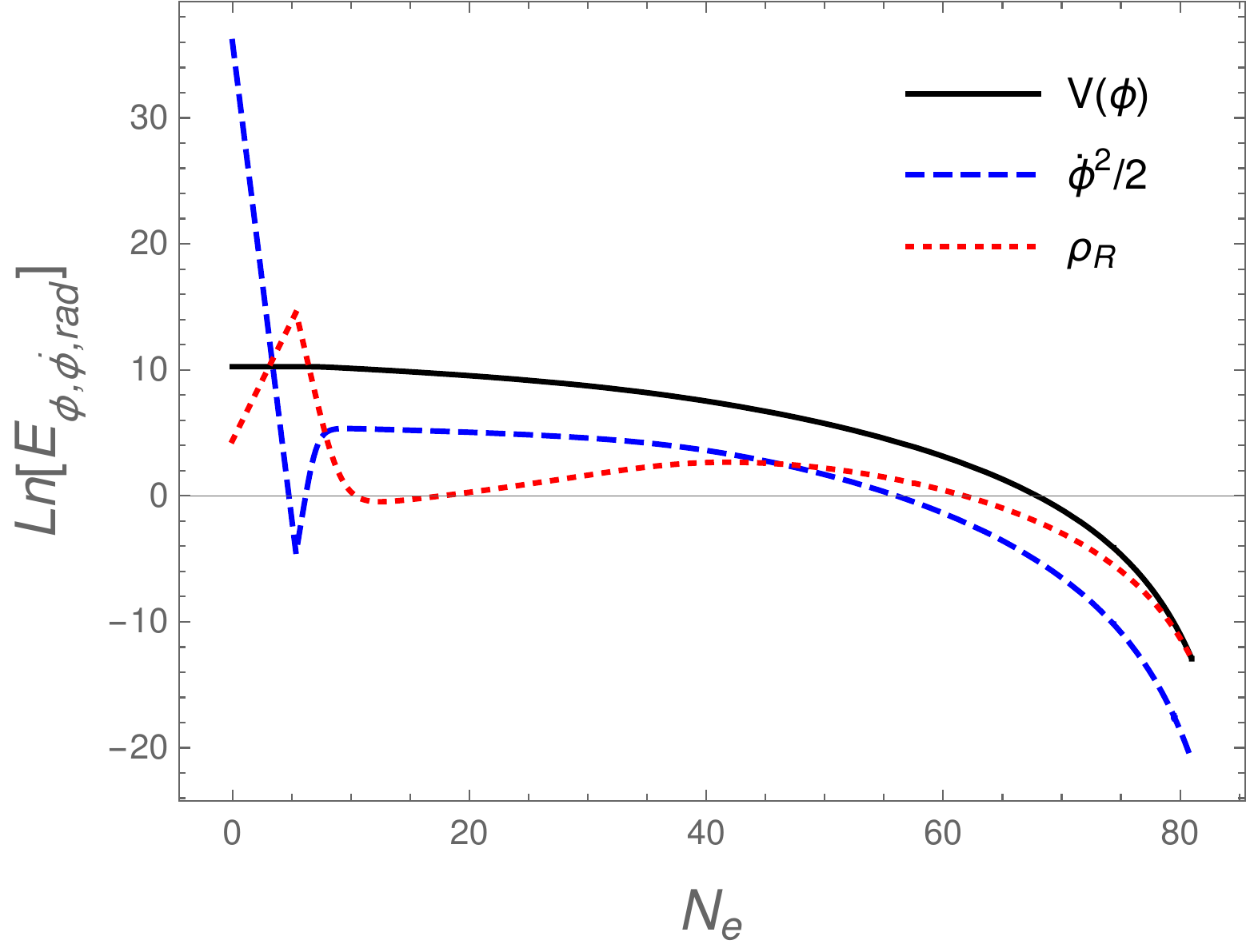}}
\subfigure[WI: linear
  dissipation]{\includegraphics[width=6cm]{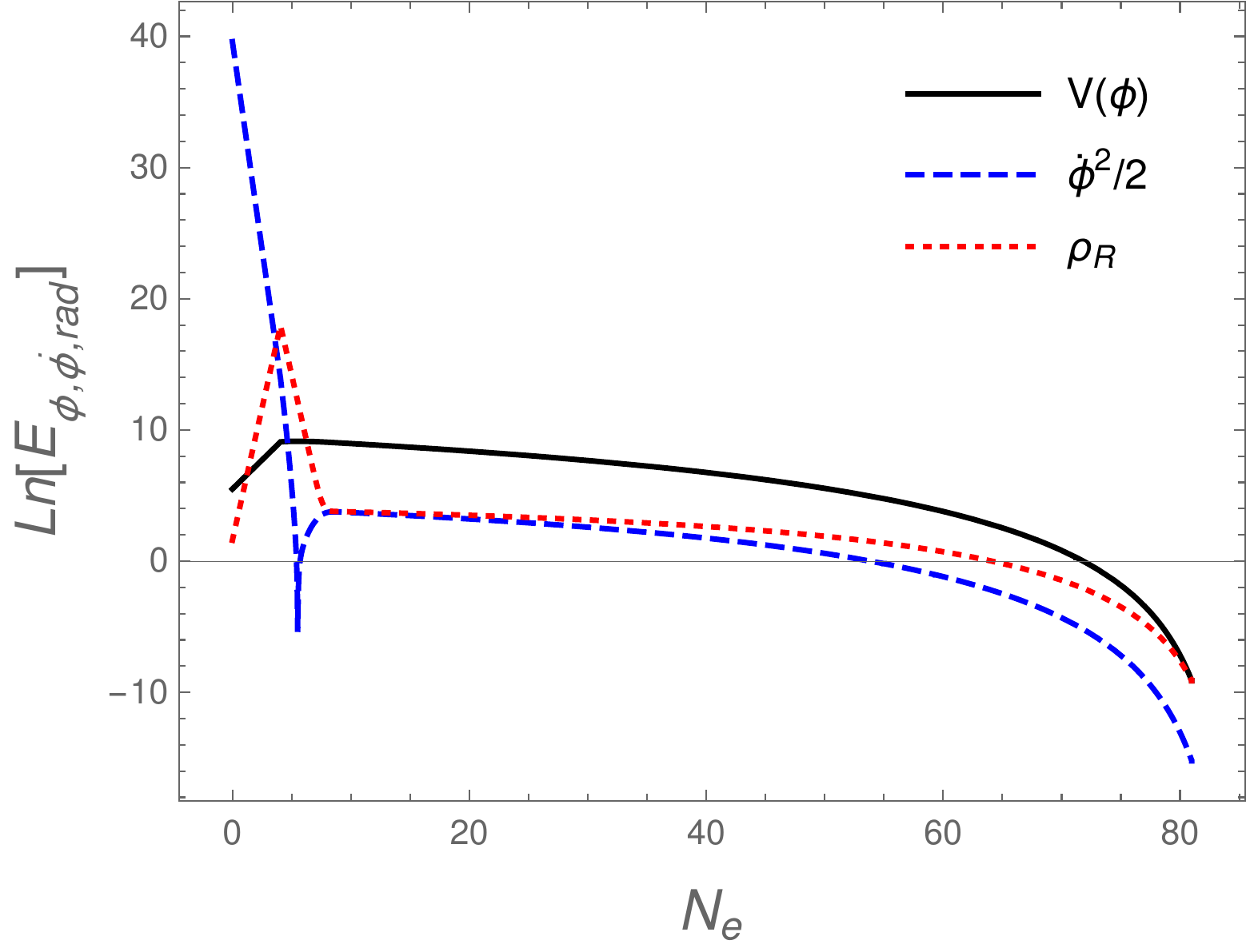}}
\caption{The evolution (in terms of number of e-folds $N_e$), from the
  bounce till the end of inflation, for the kinetic, potential and
  radiation energy densities. In the vertical axis, we have the
  natural logarithm of the energies densities, with
  $E_i=V(\phi)/\Lambda,\dot \phi^2/(2\Lambda), \rho_R/\Lambda$
  denoting  the dimensionless energy density components.  }
\label{energies}
\end{figure}
\end{center}

\begin{center}
\begin{figure}[!htb]
\subfigure[CI]{\includegraphics[width=6cm]{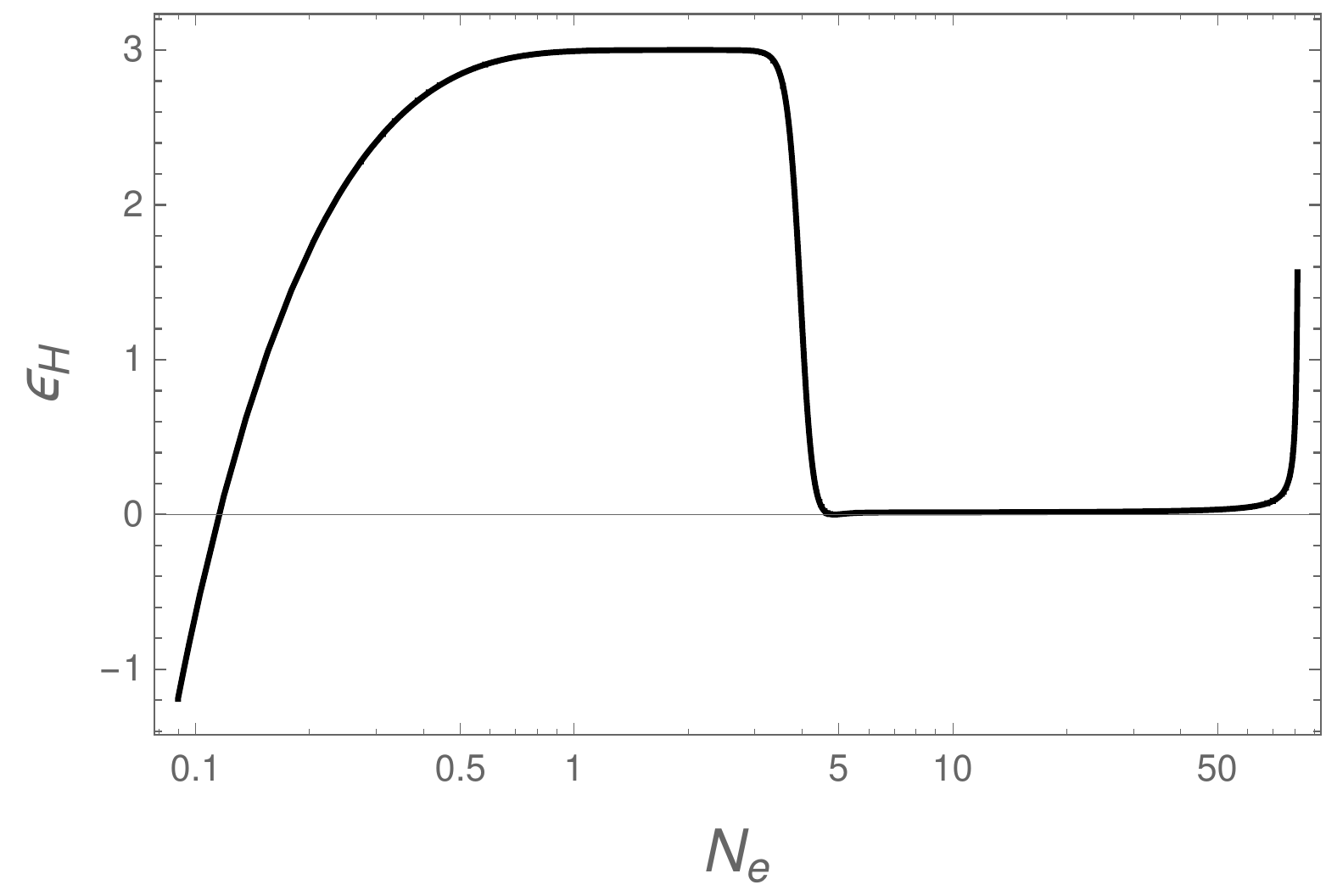}}
\subfigure[WI: cubic
  dissipation]{\includegraphics[width=6cm]{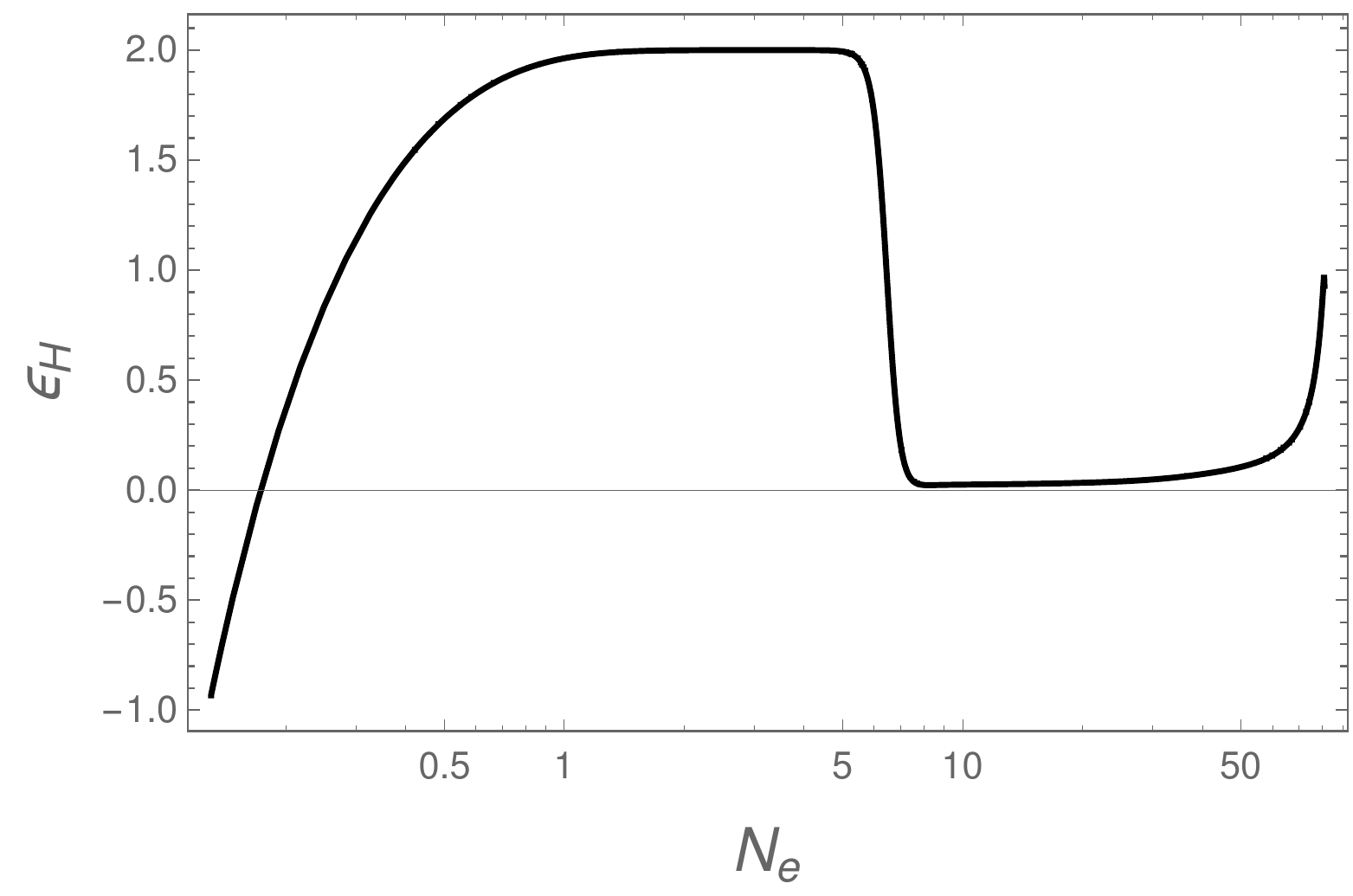}}
\subfigure[WI:linear
  dissipation]{\includegraphics[width=6cm]{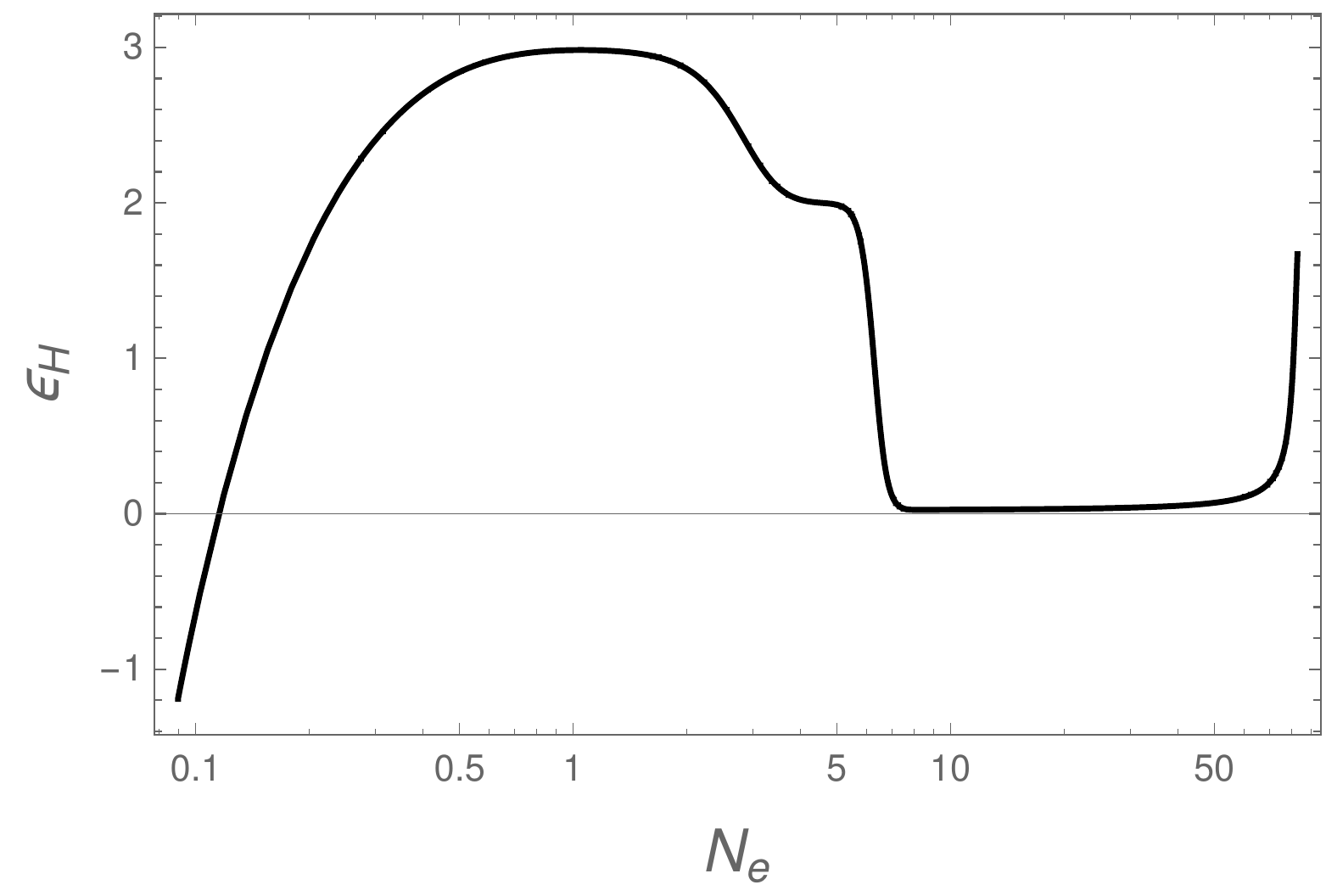}}
\caption{The evolution (in terms of number of e-folds $N_e$), from an
  instant  following the bounce till the end of inflation, for the
  slow-roll coefficient $\epsilon_H$.}
\label{epsilons}
\end{figure}
\end{center}

In {}Figs.~\ref{energies} and \ref{epsilons} we illustrate the
dynamics of WI and we also contrast it with the case of CI. {}For
convenience, in the WI case, we use the models for which the
dissipation coefficients correspond to the central values in
{}Fig.~\ref{fignsXr}, i.e., for $Q_* = 6.7 \times 10^{-3}$ in the
cubic dissipation coefficient case and for $Q_* = 0.92$ in the linear
dissipation one.

In {}Fig.~\ref{energies} it is shown the evolution of each of the
energy density components, i.e., the potential energy density
$V(\phi)$, the kinetic energy density $\dot \phi^2/2$ and the
radiation energy density $\rho_R$. As already clarified before,
our initial conditions at the bounce (where we set $N_e=0$  in the
numerical calculations)  always start with a kinetic energy dominated
regime, with $\dot \phi_B^2/2 \gg V(\phi_B) \gg \rho_{R,B}$. 
Notice that this refers to a  stiff like matter initial
evolution (with equation of state $w=1$). The kinetic energy quickly
dilutes away, since $\rho_{\dot \phi} \propto 1/a^6$. Then, around
$N_e\sim 5$, the potential energy density starts to dominate and the
inflationary regime starts. Note that in the CI case, shown in the
panel (a) in {}Fig.~\ref{energies}, inflation ends when the kinetic
energy tends to the potential energy value and the usual (p)reheating
phase can follow subsequently.  The situation is different in the two
WI cases. As shown in the panels (b) and (c) in {}Fig.~\ref{energies},
inflation tends to end now when the inflaton potential energy density reaches
an equality with the radiation energy density produced by the
dissipative process in WI. As it is typical in WI,  inflation ends
entering smoothly into the radiation dominated regime, without a subsequent
(p)reheating phase a priori.  In the two panels for the WI examples,
we also see that immediately after the bounce, the radiation density
is still very small relative to the potential energy. This phase
actually corresponds to a superinflation phase (happening also in the
CI case), which is a typical behavior seen for LQC in the presence of
scalar fields (see, e.g.,
Refs.~\cite{Ashtekar:2009mm,Ashtekar:2011rm}). This super inflation
regime starts right after the bounce, when $H=0$ and lasts till $H$
reaches its maximum values ($\dot H=0$).

Note also that in the WI case, the fast evolution of the inflaton
field also triggers a large  radiation production through the
dissipative term. The radiation energy density now raises  above the
potential energy and this coincides with the end of the
superinflation. {}From this point on, the system behaves as radiation
dominated and the radiation energy density decreases as $\rho_R
\propto 1/a^4$ and faster than the potential energy density, till the
latter dominates (at around $N_e \sim 8$ in the two WI cases shown) and
the regular inflationary regime starts.

The evolution behavior in the three regimes: superinflation
immediately following the bounce, a noinflationary regime and the
inflationary regime, can also be seen explicitly in
{}Fig.~\ref{epsilons}, where we show the slow-roll coefficient
\begin{equation}
\epsilon_H = - \frac{\dot H}{H^2},
\label{epsH}
\end{equation}
an instant after the bounce.  It can be clearly seen in all three
panels of {}Fig.~\ref{epsilons} that we have a very short
superinflation accelerated phase (with $\epsilon_H <0$)  lasting a
fraction of an e-fold, which, after a very short transient phase, it
is followed by a regular non-accelerated evolution (with $\epsilon_H >
1$)  that lasts around 4 to 8 e-folds. This phase is then followed  by
the regular inflationary phase lasting around the remaining 73 to 77
or so e-folds.

\section{Probability of WI in LQC}
\label{probab}

The  equation of motion for the volume variable $v(t)$ of LQC (defined by Eq.~(\ref{eqv})) 
and for the inflaton $\phi(t)$ can
be written in WI as
\begin{eqnarray}
\label{eomv}
\ddot{v} &=& \frac{3 v}{ M_{\rm Pl}^2} \left[ \rho
  \left(1-\frac{\rho}{\rho_{\rm cr}} \right) \right.  \nonumber \\ &-&
  \left. \frac{1}{2} \left( 1- 2\frac{\rho}{\rho_{\rm cr}}
  \right)\left(\dot\phi^2+ \frac{4}{3}\rho_R\right) \right],
\end{eqnarray}
and
\begin{equation}\label{eomphi}
\ddot{\phi} + \frac{\dot{v}}{v}\dot{\phi} +\Upsilon \dot\phi +
V_{,\phi} = 0,
\end{equation}
with $\rho \equiv \dot \phi^2/2 + V(\phi) + \rho_R$ in
Eq.~(\ref{eomv}).

Here we will follow closely the procedure of
Ref.~\cite{Ashtekar:2011rm} in their definition of the Liouville
measure used in the  definition of the a priori probability of
inflation and extend it to WI.  Let us denote the phase space as
$\Gamma$. It consists of quadruplets $(v, b; \phi, p_{\phi})$, with
$\lambda b \in [0, \pi/2]$ and $p_{\phi}=2\pi \gamma l_{\rm Pl}^{2} v
\dot{\phi}$, where $l_{\rm Pl} \equiv 1/(\sqrt{8 \pi} M_{\rm Pl})$ is
the Planck length.  The Liouville measure on $\Gamma$ is simply
$d\mu_{L} = dv \; db \; d\phi \; dp_{\phi}$.  The LQC Friedmann
equation implies that these variables must lie on a constraint surface
$\bar{\Gamma}$ defined by

\begin{equation}\label{friedmann}
\frac{3\pi}{2\lambda^{2}} \sin^{2} \lambda b = 8\pi M_{Pl}^{2}
\frac{p_{\phi}^{2}}{2v^{2}} + \frac{\pi
  \gamma^{2}}{2M_{Pl}^{2}}\left[V(\phi)+\rho_R\right].
\end{equation}
The variables $(v, b; \phi, p_{\phi})$ evolve via the system of
coupled equations,
\begin{align}
&\dot{v} = \frac{3v}{2\gamma} \frac{\sin 2\lambda b}{\lambda},  \\ &
  \dot{b}=-8M_{Pl}^{2} \frac{ p_{\phi}^{2}}{ \gamma v^{2}}, \\  &
  \dot{\phi} = 4M_{Pl}^{2}\frac{p_{\phi}}{\gamma v},
  \\ &\dot{p}_{\phi} = -\frac{ \gamma |v|}{4M_{Pl}^{2}} \left(
  V_{,\phi} + \Upsilon \dot{\phi} \right), \\& \dot{\rho}_{\rm R}=
  -\frac{4}{3}\frac{\dot{v}}{v}\rho_R + \Upsilon \dot{\phi}^{2}. 
\end{align}
We can see that the production of radiation during inflation implies
in an additional term in the expression of $\dot{p}_{\phi}$.

By using the above equations we can analyze which fraction of all
possible initial conditions evolves to an inflationary stage with the
expected amount of slow-roll expansion. This will give us the
probability that a sufficient amount of inflation occurs, which will
provide us with means to address the question of the naturalness of
inflation in the more general context of a dissipative system. Below
we will then compute this probability for the two  models of WI
discussed in the previous sections and for each of the values of
dissipation coefficient ratio (models) given in Tab.~\ref{tab:models}.

In order to compute the probability of WI in LQC we need
to obtain the Liouville measure on the space $S$ of solutions to the
equations of motion (\ref{eomv}) and (\ref{eomphi}).  This space of
solutions $S$ is found to be isomorphic to a gauge fixed surface, a
surface $\hat{\Gamma}$ of $\bar{\Gamma}$ which is intersected by each
dynamical trajectory only once~\cite{Ashtekar:2011rm}. Since the LQC
variable $b$ is monotonic along dynamical trajectories, we can choose
$b =b_{0}$ (where $b_{0}$ is a fixed constant) within
$\bar{\Gamma}$. The Liouville measure   $d\hat{\mu}_{L}$ on $S$ is
independent of the choice of $b_{0}$. The most natural choice in LQC
is to set $b_{0}=\pi/2\lambda$, the 'bounce surface'.   The induced
measure on $S$, $d\hat{\mu}_{L} = (p_{\phi}/v)d\phi dv $ can then be
written, using Eq.~(\ref{friedmann}), as,
\begin{eqnarray}\label{measure}
d\hat{\mu}&=& \frac{1}{\sqrt{8\pi}M_{Pl}}\left\{ \frac{3\pi}{\lambda^{2}}
  \sin^{2}( \lambda b_{0}) 
\right. \nonumber \\
&-& \left.  \frac{\pi
    \gamma^{2}}{M_{Pl}^{2}}\left[V(\phi)+\rho_R\right] \right\}^{1/2}d\phi dv.  
\end{eqnarray}
Now, if $(\phi(t), v(t))$ is a solution to the equations of the system and let
$\alpha$ be some constant, then $(\phi(t), \alpha v(t))$ is also a
solution, since the rescaling by a constant  corresponds to a
rescaling of spatial coordinates (or of the fiducial cell) under which
physics does not change. Therefore, this transformation corresponds to
a gauge motion~\cite{Ashtekar:2011rm}.  When computing the
probabilities we can fix $v=v_{0}$ in the integrals and we can define
the volume $v_{0}$ at the bounce to be 1. $v_{0}$  rescales the
measure by a constant and therefore drops out in the calculations of
probabilities. In conclusion, we made two gauge choices, $b=b_{0}$ and
$v=v_{0}$.

\begin{table}[!ht]
\caption{The  probabilities for each of the models represented by the points
shown in {}Fig.~\ref{fignsXr}. The value for CI is also shown for comparison.}
\begin{center}
\begin{tabular}{c|c|c|c}
  \hline  \hline 
  ${\rm Case}$ & $\Upsilon$ &  $Q_*$  & $P(E_{N_e > 81})$  
  \\ 
\hline 
CI  & -   & -  &  0.99696 
\\ 
\hline
  &   &  $1.8 \times 10^{-5}$ & $0.99756$ 
\\   
WI &  $\propto\frac{T^3}{\phi^2}$  &  $6.7 \times 10^{-3}$ & $0.99819 $  
\\    
  &   & $0.053$ & $0.998438 $ 
\\ 
\hline 
   &  &  $4.0 \times 10^{-7}$ & $0.99758$
\\  
WI  & $\propto T$  &  $0.92$ & $0.99944$ 
\\    
  &  &  $3.08$ & $0.99969$ 
\\    
\hline
\end{tabular}
\end{center}
\label{tab:prob} 
\end{table}

The probability of occurrence of an event $E$ can then be written as the
fractional volume of the region $R(E)$ in $S$ associated with  the
solutions in which $E$ occurs. We can define this probability as
\begin{equation}
P(E)=\frac{\int_{R(E)} d\hat{\mu}_{L}}{\int_{S} d\hat{\mu}_{L}}.
\end{equation}
Therefore, we can consider our gauge choices in Eq.~(\ref{measure}) in order to
write the  probability for the occurrence of a determined amount of
slow-roll inflation as~\cite{Ashtekar:2011rm}

\begin{equation} \label{P(E)}
\!\!\!P(E) \!=\! \frac{1}{N}   \int_{I(E)} \!
\sqrt{\frac{\gamma^2 \rho_{\rm cr}}{8 M_{Pl}^{6}}} \left[1 -  \frac{V(\phi)+\rho_R}{\rho_{\rm cr}}
  \right]^{1/2} \! d\phi ,
\end{equation}
where we have used Eq.~(\ref{rhocr}) and in the integration sign of the above equation $I(E)$ indicates 
that the integral limits correspond to the
limits on $\phi$ that generates the determined amount of slow roll
inflation. The normalization $N$ in Eq.~(\ref{P(E)}) is given by 
\begin{equation}
N =    \int_{-\phi_{B}}^{+\phi_B}\sqrt{\frac{\gamma^2 \rho_{\rm cr}}{8 M_{Pl}^{6}}} 
\left[1 -  \frac{V(\phi)+\rho_R}{\rho_{\rm cr}}
  \right]^{1/2} \! d\phi .
\end{equation}
Unlike the integral limits in  Eq.~(\ref{P(E)}), in the expression for
$N$ the integral  covers the whole range of possible initial values
for the variable $\phi$. The  quantities in the above expressions are
evaluated at the bounce.

{}For simplicity, firstly we are going to compute  the probability for 
the  occurrence of less than 81 e-folds of expansion from the bounce until 
the end of inflation, $P_{N_{e}\leq 81}$.  
The values $\phi_{\rm max}$ and $\phi_{\rm min}$,  given explicitly in Tab.~\ref{tab:models},
 correspond to the maximum and minimum values, respectively,  for the
inflaton that gives less than 81 e-folds of expansion. After computing $P_{N_{e} \leq 81}$,  the probability for 
occurrence of the expected amount of expansion (i.e., more than 81 e-folds until the end of inflation) can be 
easily obtained as $P_{N_{e}>81} = 1- P_{N_{e}\leq 81}$.   
In terms of the quartic potential for the inflaton, Eq.~(\ref{Vphi}),
we  have that the probability $P_{N_{e}\leq 81}$ can be explicitly expressed as

\begin{equation}\label{integralprobab}
\!\!\!\!\!P_{N_{e}\leq 81} \!=\! \frac{\int_{\phi_{\rm min}}^{\phi_{\rm max}} \!\! \sqrt{ 1 \!-\!  
      \frac{\sqrt{3} \gamma^3}{6} \frac{\Lambda}{M_{\rm Pl}^4} \left[
    \frac{1}{4}\left(\frac{\phi}{M_{\rm Pl}}\right)^4 \!+\! \frac{\rho_R}{\Lambda}
    \right] } d \phi }{\int_{-\phi_B}^{+\phi_B}  \!\! \sqrt{ 1 \!-\!  
      \frac{\sqrt{3} \gamma^3}{6} \frac{\Lambda}{M_{\rm Pl}^4} \left[
    \frac{1}{4}\left(\frac{\phi}{M_{\rm Pl}}\right)^4 \!+\! \frac{\rho_R}{\Lambda}
    \right] } d \phi}.
\end{equation}
Using the values given in Tab.~\ref{tab:models}, the computation of the quantity  
$P_{N_{e}>81} = 1- P_{N_{e}\leq 81}$ is straightforward.  Table~\ref{tab:prob} shows the results 
for the probability $P_{N_{e}> 81}$ of occurrence of the expected amount of inflation for all the
inflationary models, represented by the points shown in {}Fig.~\ref{fignsXr}.  
We can see that the probability
is always very close to one in all the cases analyzed. We observe that
the probability for the occurrence of a sufficient amount of WI is
larger (although slightly) than the one for CI. We also see that in
each of the WI models considered, the probability tends to increase
with the dissipation. This can be interpreted by the fact that in the
presence of dissipation, the trajectories are attracted faster to the
inflationary region and they stay longer
there~\cite{deOliveira:1997jt,Ramos:2001zw}. Thus, the probability
tends always to be larger in WI than in CI.

\section{Conclusions}
\label{concl}

In this work we have considered the warm inflationary scenario in the
context of LQC. We have analyzed the modifications in the dynamical
system  due to quantum effects on the geometry from LQC, which
produces a bounce in the place of the singularity of GR. The presence
of the bounce sets the appropriate conditions where a Liouville
measure can be defined properly and probabilities be computed.  We
have then taken advantage of this to extend previous analysis of the a
priori probability in the CI picture to the case of WI.  We have
focused our analysis for the case of a monomial quartic potential for
the inflaton, which despite being excluded in the CI case, it is not
in WI.   We have then obtained the appropriate conditions for which
the consistency with the Planck data is assured. {}From these, we were
able to determine the initial conditions set at the bounce which would
lead to trajectories in the phase space exactly passing through the
allowed Planck region in the $(r,n_s)$ plane.  Our results have
demonstrate that in WI we have a superinflation phase, as expected in
general, followed by a noninflationary radiation dominated region,
followed then by the usual slow-roll inflationary region, lasting long
enough. In the examples given, we have also shown that this
inflationary region can end smoothly in a radiation dominated regime,
as is expected in general in WI.

Our results for the probability for the occurrence of at least 81
e-folds of inflation in the WI models were all consistently higher
that in the CI picture.  {}Furthermore, with the monomial quartic inflaton
potential used and for the different  dissipation coefficients
considered,  these favorable trajectories are all consistent with the
Planck data. This is opposite to what happens in the CI case. In CI, despite the
probability of trajectories with enough e-folds of inflation is highly
probable, none of them is able to satisfy the observations.

It would be interesting, as a future work, to also study other
different inflationary potentials for WI in the LQC context. In
addition, a close  connection with model building in WI, in the lines
of e.g. Ref.~\cite{Bastero-Gil:2016qru}, would also help to access
these probabilities  in terms of the microscopic parameters (i.e., the
couplings and energy scales) of the model. This could help in
constraining further these parameters and helping in the model
building for WI.

\section*{Acknowledgments}

L.L.G. is supported by a posdoc grant "P\'os-Doutorado Nota 10" from
Funda\c{c}\~ao Carlos Chagas Filho de Amparo \'a Pesquisa do Estado do
Rio de Janeiro (FAPERJ), No. E - 26/202.511/2017.  R.O.R is partially
supported by research grants from Conselho Nacional de Desenvolvimento
Cient\'{\i}fico e Tecnol\'ogico (CNPq), grant No. 302545/2017-4 and
Funda\c{c}\~ao Carlos Chagas Filho de Amparo \'a Pesquisa do Estado do
Rio de Janeiro (FAPERJ), grant No.  E - 26/202.892/2017. 


\end{document}